\documentclass[12pt,preprint]{aastex}
\newcommand{\htwo}{H\,{\scriptsize\sc{II}} }

\begin{document}

 \title{The Extremely High-Velocity Outflow from the Luminous Young Stellar Object G5.89$-$0.39}

 \author{Yu-Nung Su,\altaffilmark{1} Sheng-Yuan Liu,\altaffilmark{1,2} Huei-Ru Chen\altaffilmark{2,1}, and Ya-Wen Tang\altaffilmark{3,4}}

 \altaffiltext{1}{Institute of Astronomy and Astrophysics, Academia Sinica, P.O. Box 23-141, Taipei 106, Taiwan; ynsu@asiaa.sinica.edu.tw}
 \altaffiltext{2}{Institute of Astronomy and Department of Physics, National Tsing Hua University, Hsinchu, Taiwan}
 \altaffiltext{3}{Universit\'{e} de Bordeaux, Observatoire Aquitain des Sciences de l$^\prime$Univers, BP 89, 33271 Floirac, France}
 \altaffiltext{4}{CNRS/INSU $-$ UMR5804, Laboratoire d$^\prime$Astrophysique de Bordeaux, BP 89, 33271 Floirac, France}

\begin{abstract}

We have imaged the extremely high-velocity outflowing gas in CO
(2$-$1) and (3$-$2) associated with the shell-like ultracompact
\htwo region G5.89$-$0.39 at a resolution of $\sim$3$\arcsec$
(corresponding to $\sim$4000 AU) with the Submillimeter Array. The
integrated high-velocity ($\gtrsim$45 km s$^{-1}$) CO emission
reveals at least three blueshifted lobes and two redshifted lobes.
These lobes belong to two outflows, one oriented N-S, the other
NW-SE. The NW-SE outflow is likely identical to the previously
detected Br$\gamma$ outflow. Furthermore, these outflow lobes all
clearly show a Hubble-like kinematic structure. For the first time,
we estimate the temperature of the outflowing gas as a function of
velocity with the large velocity gradient calculations. Our results
reveal a clear increasing trend of temperature with gas velocity.
The observational features of the extremely high-velocity gas
associated with G5.89$-$0.39 qualitatively favor the jet-driven bow
shock model.

\end{abstract}

\section{Introduction \label{s-intro}}

Although molecular outflows have been commonly identified around
both low- and high-mass young stellar objects (YSOs)
\citep[e.g.,][]{bac99,arc07}, it is not clear how the bulk of the
outflowing gas is accelerated. Past observations show that molecular
outflows can often be divided into two components --- the
``classical'' less-collimated component with low-velocity and the
highly collimated component with extremely high-velocity (50$-$150
km s$^{-1}$, hereafter EHV) \citep{bac99,hir06,pal06}. Exploring the
physical conditions of the EHV molecular gas in details will be
helpful for clarifying its role in star formation processes.

The massive star forming region G5.89$-$0.39 (hereafter G5.89) is
associated with energetic CO outflows with velocities up to
$\sim$$\pm$70 km s$^{-1}$ from its systemic velocity, among the
highest gas velocities that have been detected in molecular outflows
\citep[e.g.,][]{cho93}. G5.89 also harbors a shell-like ultracompact
(UC) \htwo region powered by a young O-type star (hereafter Feldt's
star) visible in the near-IR \citep[][]{wc89,fel03}. Based on the
deduced gas temperature structure, \citet{su09} conclude that
Feldt's star supplies the majority of energy for heating the dense
gas enshrouding the UC \htwo region.

The outflowing gas associated with G5.89 was first identified by
\citet{har88} in CO (1$-$0) with the IRAM 30-m telescope. Subsequent
studies reported outflows with notably different orientations in
various tracers. For example, SiO (5$-$4) observations with the
Submillimeter Array\footnote{The Submillimeter Array is a joint
project between the Smithsonian Astrophysical Observatory and the
Academia Sinica Institute of Astronomy and Astrophysics, and is
funded by the Smithsonian Institution and the Academia Sinica.}
(SMA) resolved a bipolar structure along a position angle (P.A.) of
$\sim$30$^\circ$ \citep{sol04}. On the other hand, sub-arcsecond CO
(3$-$2) images made with the SMA  \citep{hun08} detected a
well-collimated N-S outflow and tentatively a compact NW-SE outflow
associated with the Br$\gamma$ outflow revealed by \citet{pug06}.
Both CO outflows demonstrate EHV gas, with velocities $\gtrsim$50 km
s$^{-1}$. Furthermore, BIMA observations in CO (1$-$0) and HCO$^+ $
(1$-$0) identified an outflow along the E-W direction \citep{wat07}.
These results imply the existence of multiple YSOs toward G5.89.
Indeed, five sub-mm dust sources (denoted as SMA1, SMA2, SMA-N,
SMA-E, and SMA-S) were located in the close vicinity of the UC \htwo
region G5.89 \citep{hun08}.

Here we present SMA observations of G5.89 in CO (2$-$1) and (3$-$2)
at resolutions of $\sim$3$\arcsec$. We perform large velocity
gradient (LVG) calculations to explore the physical conditions of
the EHV outflowing gas. We then discuss the nature of the EHV
components and their connection to the gas accelerating processes.

\section{Observations and Data Reduction}

The observations were carried out with the SMA on 2008 April 17 and
May 2 in the band of 230 GHz, and on 2006 July 27 and September 10
in the band of 345 GHz. The phase center was R.A. =
18$^h$00$^m$30.32$^s$ (J2000) and decl. =
$-$24\degr04\arcmin00.50$\arcsec$ (J2000). At 230 GHz band, CO
(2$-$1) was observed with seven antennas in the compact
configuration, which resulted in projected baselines ranging from
about 9~m to 120~m (7 to 90 k$\lambda$). For the CO (3$-$2)
observations at 345 GHz band with also seven antennas in the compact
configuration, the projected baselines ranged from about 9~m to 70~m
(10 to 80 k$\lambda$). The half-power width of the SMA primary beam
was $\sim$51$\arcsec$ at 230 GHz and $\sim$34$\arcsec$ at 345 GHz.
The total available double-sideband bandwidth was 4 GHz. The
spectral resolutions of 230 GHz and 345 GHz observations were 0.41
MHz (0.53 km s$^{-1}$) and 0.81 MHz (0.71 km s$^{-1}$),
respectively. See \citet{ho04} for more complete specifications of
the SMA.

For the 345 GHz band observations, the flux calibrator was Uranus
and the bandpass calibrators were Uranus (July 27 and September 10),
Callisto (July 27), and 3C279 (September 10). The nearby compact
radio sources 1626$-$298 (\emph{S} $\sim$ 2.2 Jy in July 27 and 1.8
Jy in September 10) and 1924$-$292 (\emph{S} $\sim$ 3.4 Jy in July
27 and 2.9 Jy in September 10) served as complex gain calibrators.
See \citet{su09} for the calibration information of the 230 GHz band
observations. For both 230 GHz and 345 GHz band observations, the
absolute flux density scales were estimated to have an uncertainty
of $\sim$20\% . We calibrated the data using the MIR software
package adapted for the SMA from the software package developed
originally for the OVRO MMA \citep{sco93}. We made maps using the
MIRIAD package \citep{sau95}. With robust weighting, the synthesized
beam size was about 3.2$\arcsec$ $\times$ 2.4$\arcsec$ at P.A. of
40.8$^\circ$ for the CO (2$-$1) maps, and about 3.0$\arcsec$
$\times$ 2.0$\arcsec$ at P.A. of 13.5$^\circ$ for the CO (3$-$2)
maps. We smoothed our data to 2.0 km s$^{-1}$ resolution for the
analysis presented below.  The rms noise level in a 2.0 km s$^{-1}$
velocity bin is $\sim$35 mJy beam$^{-1}$ ($\sim$0.10 K) at 230 GHz
and $\sim$90 mJy beam$^{-1}$ ($\sim$0.15 K) at 345 GHz.

\section{CO (2$-$1) \& (3$-$2) Emissions \label{resultco}}

\begin{figure}
 \epsscale{.8}
 \plotone{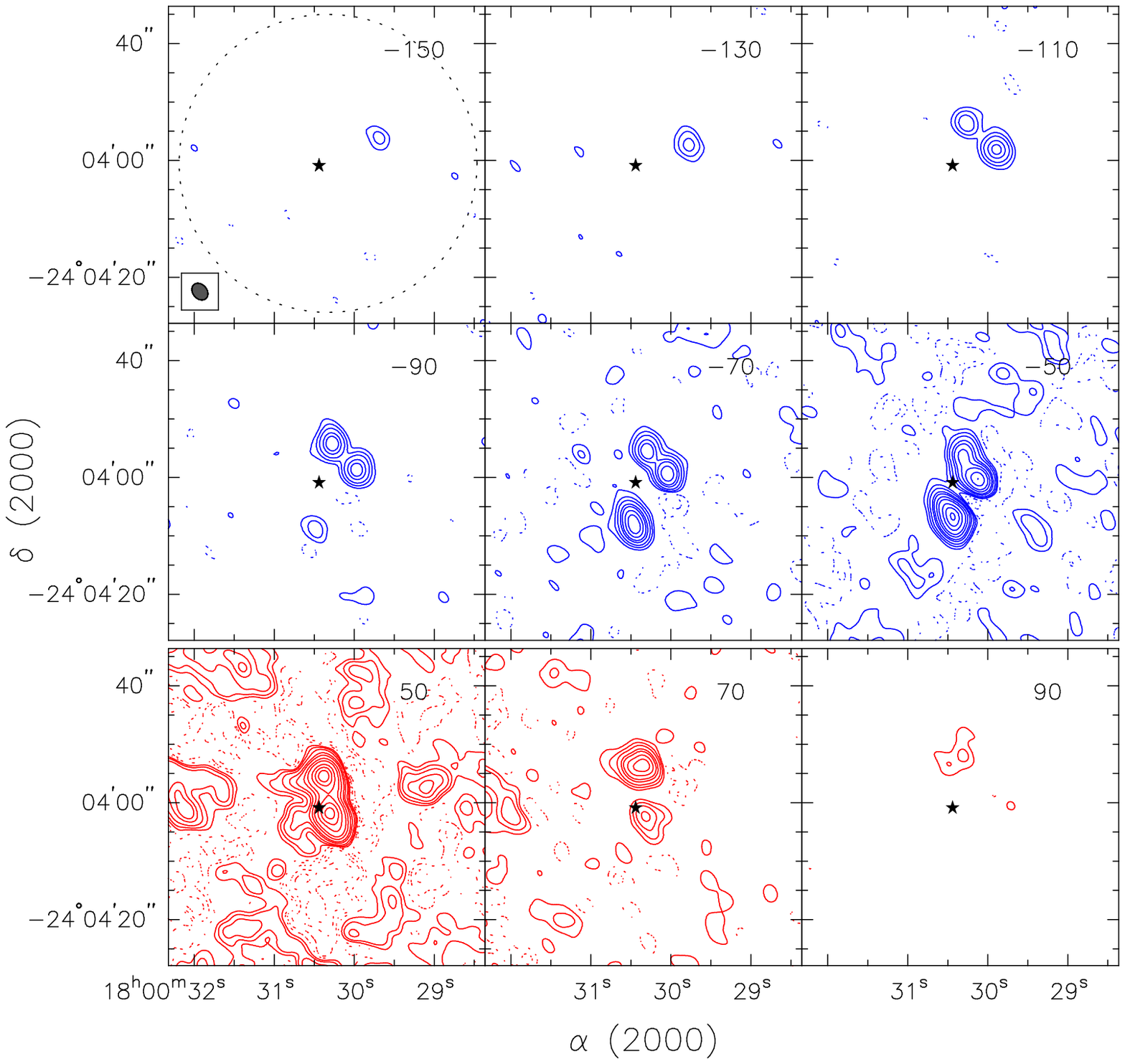}
 \caption{Channel maps of G5.89$-$0.39 in CO (2$-$1) emission at high velocity. The local standard of rest
 velocity of each channel's center is indicated in the upper
 right corner of each panel. The LSR velocity of the system is
 about $+$9.0 km s$^{-1}$. Solid contours are at 4, 10, 20, 40, 60,
 100, 150, 250, 350, and 450 $\times$ $\sigma$, and dotted contours
 indicate $-$4, $-$10, $-$20, $-$40, and $-$60 $\times$ $\sigma$, where $\sigma$ =
 11 mJy beam$^{-1}$. The dotted circle shown in the first panel
 represents the primary beam of the SMA observations. The synthesized
 beam is shown in the lower left corner of the first panel. The
 asterisk represents the position of the \htwo region powering star, i.e., Feldt's star \citep{fel03}. \label{co21chan}}
\end{figure}

Figure~\ref{co21chan} shows channel maps of the blueshifted and
redshifted CO (2$-$1) emissions associated with G5.89. For display
purposes, we have smoothed the channel maps to a velocity resolution
of 20.0 km s$^{-1}$. The spatial distributions of CO (2$-$1) and
(3$-$2) agree with each other very well, and the velocity extents of
the CO (2$-$1) and (3$-$2) outflows are comparable. Emissions are
detected to velocities $v_{flow}$ $\approx$ 80 km s$^{-1}$ in the
redshifted lobes and $v_{flow}$ $\approx$ 160 km s$^{-1}$ in the
blueshifted lobes, where $v_{flow}$ $\equiv$ $|V_{flow}-V_{lsr}|$
with $V_{flow}$ being the apparent outflow velocity and $V_{lsr}$ (=
$+$9.0 km s$^{-1}$) the systemic velocity of G5.89. The EHV line
wing detected here is significantly broader than that reported by
single-dish observations \citep[e.g.,][]{cho93}. The difference is
due to the relatively poor sensitivities of single-dish
observations. The structure of the CO gas with $v_{flow}$ $\lesssim$
40 km s$^{-1}$ is extended and complex. Due to the lack of
short-spacing data to recover the extended structure filtered out by
our SMA observations, in Figure \ref{co21chan} the low-velocity
components are excluded.

The \emph{left} and \emph{right} panels of Figure \ref{co21mn0} show
the integrated blueshifted and redshifted emissions of CO (2$-$1)
and (3$-$2), respectively. For both transitions, the emissions are
integrated over the line wing, with 50 $\lesssim$ $v_{flow}$
$\lesssim$ 162 km s$^{-1}$ for the blueshifted gas and 42 $\lesssim$
$v_{flow}$ $\lesssim$ 80 km s$^{-1}$ for the redshifted gas. The
outflow morphologies seen in CO (2$-$1) and (3$-$2) are very
similar. Although only EHV line wing is integrated, the morphology
of the EHV gas is complicated
--- at least three bluedshifted lobes (denoted as B-N lobe, B-NW
lobe, and B-S lobe) and two redshifted lobes (denoted as R-N lobe
and R-S lobe) are revealed in the vicinity of the UC \htwo region
G5.89. There appears to be an additional redshifted lobe located
$\sim$4\arcsec~ southeast to Feldt's star.

Both B-S lobe and R-N lobe have been identified by previous SMA CO
(3$-$2) observations at sub-arcsecond resolution, and were proposed
to be a pair of the N-S bipolar outflow driven by the sub-mm dust
source SMA1 \citep{hun08}. The authors also proposed a second CO
outflow along P.A. about $-$55$^\circ$ associated with the
Br$\gamma$ outflow reported by \citet{pug06}. Our observations
confirm the existence of this NW-SE outflow, consisting of the B-NW
lobe and R-S lobe. The B-N lobe was also identified by
\citet{hun08}, although its driving source is less clear. The
existence of the B-N lobe could be explained by the N-S outflow
being close to the plane of the sky, but this is very unlikely from
the kinematic viewpoint. After correcting for projection, the N-S
outflow, if in the plane of sky, will have molecular gas with
velocity up to $>$500 km s$^{-1}$, about a factor of 3 higher than
the highest gas velocities that ever identified in molecular
outflows. Figure \ref{copv} shows the position-velocity (P-V)
diagrams of the CO (2$-$1) and (3$-$2) along P.A. 175$^\circ$
(\emph{left} panel) centered on the dust source SMA1 and 130$^\circ$
(\emph{right} panel) centered on the dust source SMA2. The
Hubble-like kinematic feature, i.e., the outflow velocity increasing
with the distance from the central YSO, is clearly discerned in both
outflows with a velocity gradient about 1000 km s$^{-1}$ pc$^{-1}$.

Assuming an inclination of 45$^\circ$ and a distance of 1.28 kpc
\citep{mot11}, the inferred dynamical timescales range from 400
years to 650 years. Since the axes of these outflows are unlikely
close to either the plane of the sky or the line of sight, the
derived dynamical timescales should be accurate to within a factor
of 3$-$4. The observed outflow lobes therefore represent the most
recent outflow activities in this region.

\section{Line Ratios \& Excitation Conditions of the Outflowing Gas \label{properties}}

Given the fairly good signal-to-noise ratio of the CO emission
detected toward G5.89 even at EHV line wing, it is feasible to
estimate the physical parameters with a relative narrow velocity bin
(10 km s$^{-1}$) and investigate their velocity-dependence.
Table~\ref{tb-ratio-b} summarizes the observed CO (2$-$1)
intensities (in brightness temperature scale) as well as the
intensity ratios of CO (3$-2$) and (2$-$1) as a function of gas
velocity for all the observed lobes.

For a proper comparison, both CO (2$-$1) and (3$-$2) maps were
reconstructed with a clean beam of 3.4$\arcsec$ in order to match
the resolutions. For each velocity bin, furthermore, we simply
report the value calculated from the CO (2$-$1) peak toward each
lobe rather than the value averaged over entire lobe to avoid the
difficulty of emission separation from various lobes, in particular,
at relatively low-velocity channels. We emphasize that for each
outflow lobe, the peak positions of CO (2$-$1) and (3$-$2) in
various velocity bins agree very well with each other, with both
median and mean offsets of $\sim$0.25$\arcsec$, and no systematic
offsets can be discerned. Consequently we do not expect any
systematic bias for the physical conditions estimated afterward. As
listed in Table \ref{tb-ratio-b}, most deduced line ratios are in
the range of 0.6$-$1.2.

To estimate the physical conditions of the outflowing gas, we
performed the LVG analysis with the code written by L. G. Mundy and
implemented as part of the MIRIAD package \citep{sau95}. We assumed
the canonical CO fractional abundance of 10$^{-4}$ and a velocity
gradient of 1000 km s$^{-1}$ pc$^{-1}$ as estimated from the
Hubble-like kinematic structure (in Figure \ref{copv}). Figure
\ref{lvg-temp} shows the gas temperature estimated from the LVG
calculations versus gas velocity. The plot exhibits a clear
increasing trend of gas temperature with outflow velocity. Such a
trend can be discerned toward all the lobes, although the trend of
the B-N lobe is less clear. In the case of B-NW lobe, there appears
to be a temperature jump at outflow velocities of $\sim$150 km
s$^{-1}$, with gas temperature ranging from 1200$-$1800 K for
\emph{v}$_{flow}$ $\gtrsim$150 km s$^{-1}$ and $\lesssim$150 K for
\emph{v}$_{flow}$ $\lesssim$150 km s$^{-1}$. The estimated
temperature of the above highest-velocity bins in the B-NW lobe will
be reduced to 400$-$500 K hence eliminating the above-mentioned
temperature jump if the assumed CO fractional abundance is lowered
by a factor of 2. The occurrence of abundance reduction may be
attributed to partial dissociation or even ionization of CO
molecules under high-excitation conditions.

A few factors may contribute to the uncertainty of the temperature
estimates reported above. Given the flux calibration uncertainty of
20\%, the gas temperature obtained with LVG calculations would vary
by a factor of 10 or so. In Figure 4, we show with the blue-shaded
area the 1-$\sigma$ temperature range when allowing flux variations
up to 20\% in the B-NW lobe. Although the temperature uncertainty
appears fairly large, we emphasize that the general increasing trend
of gas temperature with outflow velocity is indeed robust.  For a
given flux calibration error, the line ratios will be affected in
the same way in all channels and render a simple temperature shift
across all velocity bins.  For the LVG calculations, variations in
input parameters also lead to variations in the derived temperature
by a similar factor and hence do not change the general increasing
trend on temperature with gas velocity.  For example, the derived
temperature will decrease by about a factor of 3 if the adopted CO
fractional abundance is reduced by half or the velocity gradient
increases twice. Although variation in velocity gradient or CO
abundance across the line wings could mimic the derived temperature
trend with velocity, the Hubble-like kinematic feature shown in
Figure \ref{copv} makes this confusion quite unlikely. A precise gas
temperature estimate will need observations with better flux
calibration. Alternatively, observations in higher transitions of CO
will also help to constrain hot gas temperature.

\section{The Origin of the EHV Components and Their Connection to the Gas Accelerating Processes}
Observations have shown that the kinetic temperature of
high-velocity molecular gas in outflows associated with YSOs can be
as high as 100$-$1000 K. For example, the inner SiO knots in HH211
jet have a temperature excess of 300$-$500 K \citep{hir06}. In the
case of extremely active molecular jets in L1448-mm, the kinetic
temperature of the EHV bullets close to the YSO is estimated to be
$\gtrsim$ 500 K \citep[][]{nis07}. Gas temperature estimated from
near-IR H$_2$ observations typically ranges from 1000$-$2500 K
\citep{ric00,dio10}. Both hot H$_2$ emission and warm SiO bullets
are thought to trace shocked molecular gas. The broad velocity
dispersions revealed from the kinematic structures of the EHV
bullets further indicate their connections to (bow) shocks
\citep{lee01,hir06,su07}.

\begin{figure}
  \epsscale{1.1}
 \plottwo{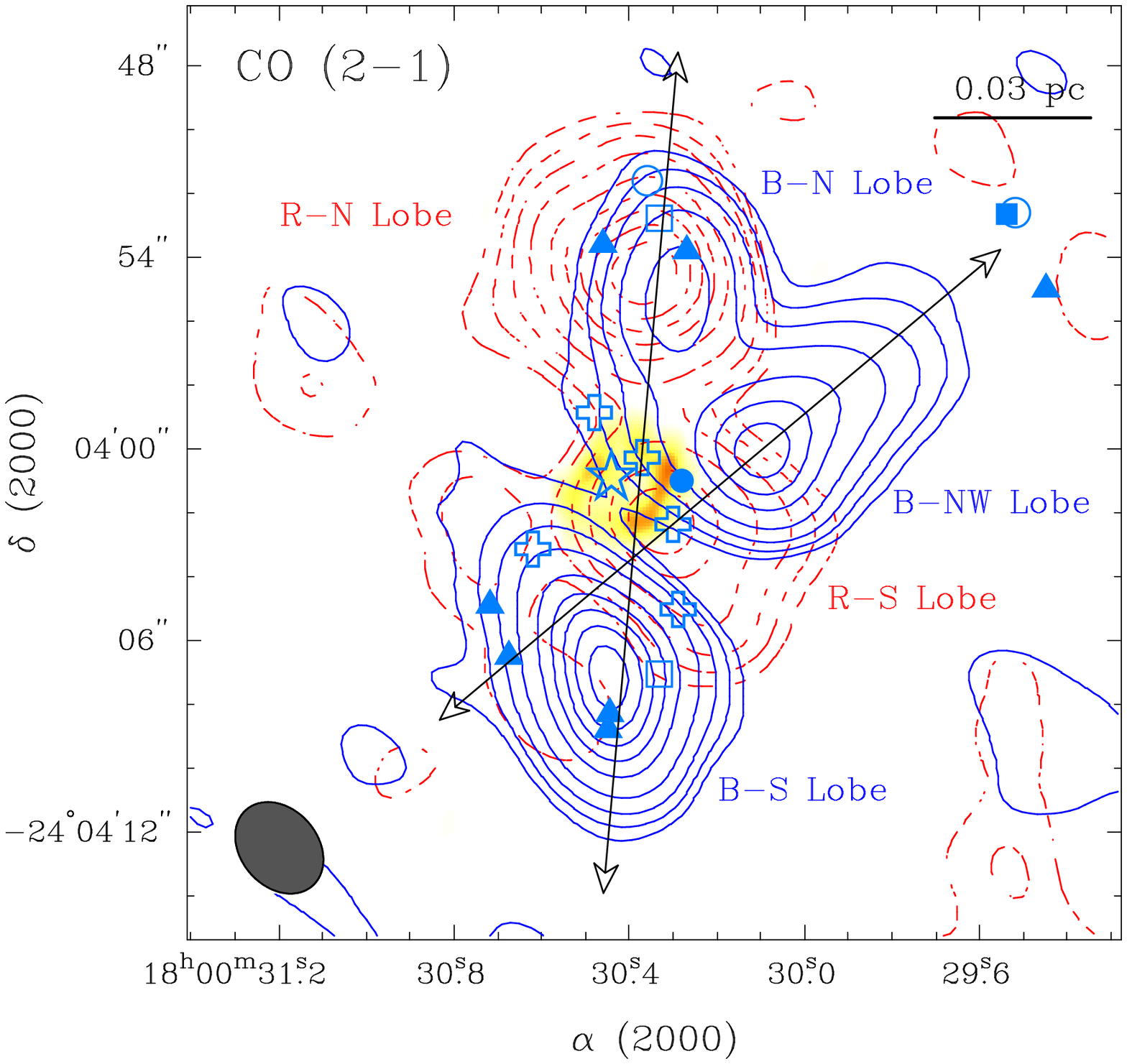}{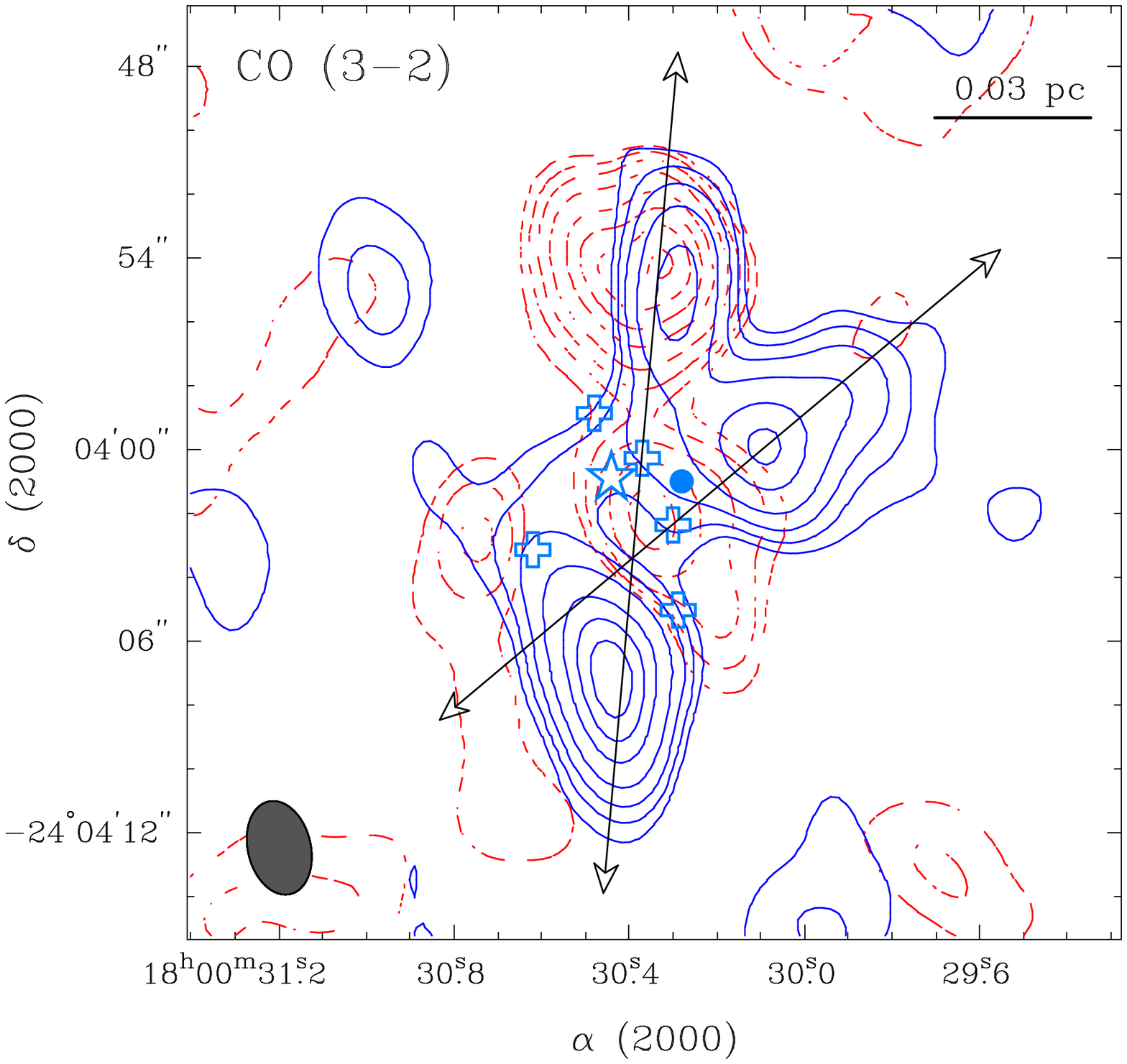}
 \caption{Extremely high-velocity molecular outflows of G5.89$-$0.39 in CO (2$-$1) (\emph{left}) and (3$-$2) (\emph{right}) imaged with the SMA.
For both transitions, the blueshifted and redshifted lobes are
integrated over the line wing, with 50 $\lesssim$ $v_{flow}$
$\lesssim$ 162 km s$^{-1}$ for the blueshifted gas and 42 $\lesssim$
$v_{flow}$ $\lesssim$ 80 km s$^{-1}$ for the redshifted gas. In the
\emph{left} panel, the overlaid color scales represent the free-free
emission at 2 cm \citep{tan09}, filled triangles are the near-IR
H$_2$ knots from \cite{pug06}, filled square is the NH$_3$ maser
from \cite{hun08}, open circles are Class I methanol maser positions
(components 1 and 2 from \citet{kur04}) and open squares are
positions of water masers (components 1 and 3 from \citet{hof96}).
In each panel, the crosses mark the position of the sub-mm dust
components reported by \citet{hun08}, the asterisk marks the
position of Feldt's star \citep{fel03}, and the filled circle labels
the Br$\gamma$ outflow center \citep{pug06}. The synthesized beam is
shown in the lower left corner. The two black lines represent the
axes of the position-velocity diagrams plotted in Figure \ref{copv}.
In the \emph{left} panel, contours are 4, 10, 20, 40, 80, 120, 160,
200, and 240 $\times$ $\sigma$, where $\sigma$ = 0.53 and 0.31 Jy
beam$^{-1}$ km s$^{-1}$, respectively, for the blue- and redshifted
components. In the \emph{right} panel, contours are 4, 10, 20, 40,
80, 120, 160, and 200 $\times$ $\sigma$, where $\sigma$ = 1.36 and
0.80 Jy beam$^{-1}$ km s$^{-1}$, respectively, for the blueshifted
and redshifted components. \label{co21mn0}}
\end{figure}

\begin{figure}
 \epsscale{0.7}
 \plotone{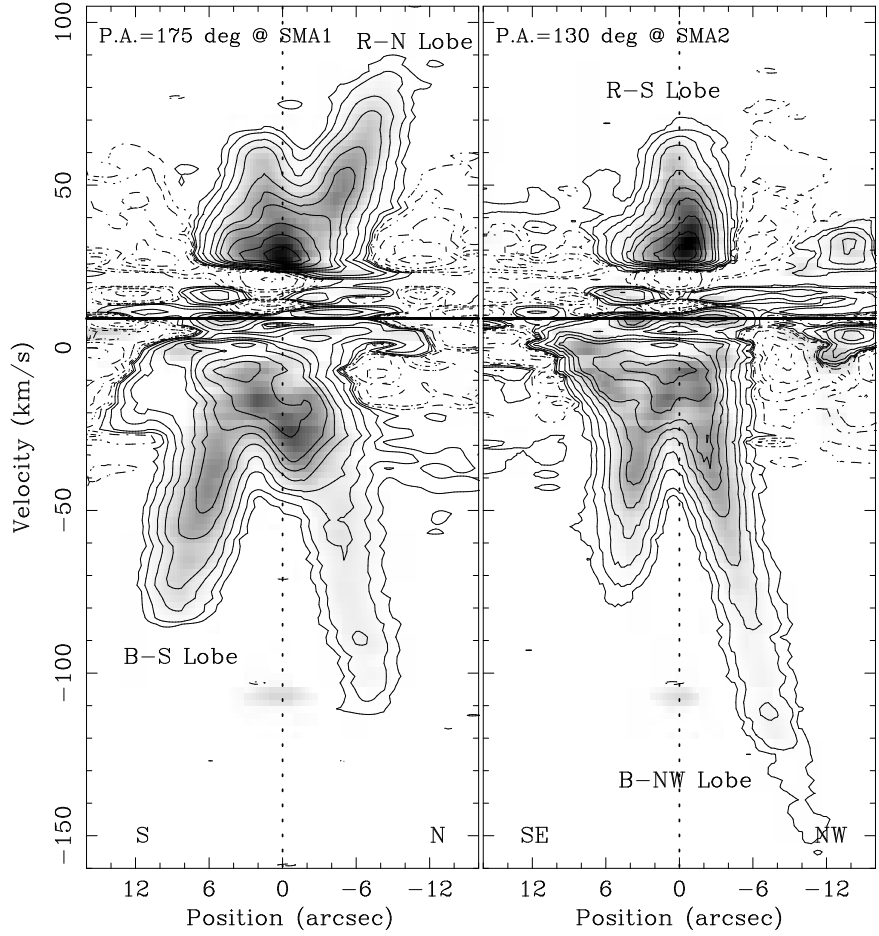}
 \caption{Position-velocity diagram of G5.89$-$0.39 in CO (2$-$1)
(\emph{contours}) and (3$-2$) (\emph{grayscales}) along the P.A. of
175$^\circ$~centered at SMA1 (\emph{left}) and P.A. of
130$^\circ$~centered at SMA2 (\emph{right}). In each panel, the
horizontal line marks the LSR velocity of the system (i.e., $+$9.0
km s$^{-1}$), and the vertical line indicates the position of SMA1
(\emph{left}) and SMA2 (\emph{right}). \label{copv}}
\end{figure}

\begin{figure}
 \epsscale{1.}
 \plotone{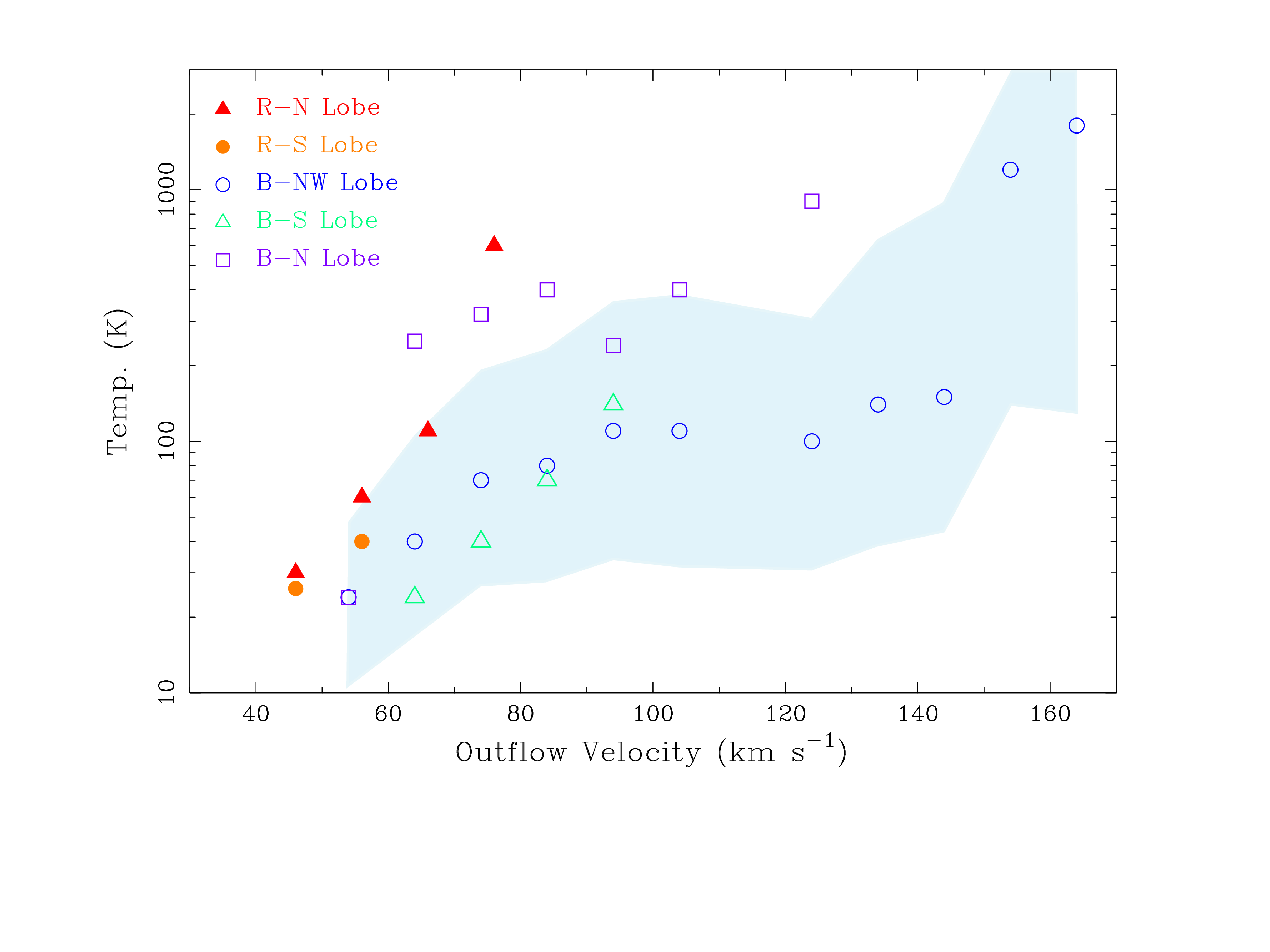}
 \caption{Temperature of the outflowing gas deduced from the LVG calculations versus the corresponding gas velocity. An increasing trend of gas
temperature with outflow velocity can be discerned toward all lobes.
The blue-shaded area represents the 1-$\sigma$ temperature
uncertainty when allowing flux variations up to 20\% in the B-NW
lobe. Note that the outflow velocity $v_{flow}$ shown here is
related to the velocity $V_{flow}$ listed in Table 1 by the
relation: $v_{flow}$ = $|V_{flow}-V_{lsr}|$ with $V_{lsr}$ = $+$9.0
km s$^{-1}$ the systemic velocity of G5.89$-$0.39. The lobes of N-S
and NW-SE outflows are represented by triangles and circles,
respectively. \label{lvg-temp}}
\end{figure}

Could the heating of the EHV gas associated with G5.89 be dominated
by shocks? Indeed, shock indicators have been detected around the
highest velocity gas of the N-S and NW-SE outflows
\citep{hof96,kur04,hun08,pug06}. For example, as shown in Figure
\ref{co21mn0} the highest-velocity gas in the R-N lobe is associated
with masers of H$_2$O and Class I CH$_3$OH as well as H$_2$ knots
(labeled as C1 and C2 by \cite{pug06}). Furthermore, the positions
of the highest velocity gas detected in the B-NW and B-S lobes
coincide with the H$_2$ knots B and group A, respectively, very
well. The highest velocity in the B-NW lobe is also associated with
masers of NH$_3$ and Class I CH$_3$OH. The hot (about 150-1800 K)
molecular gas in the highest-velocity bins of the N-S and NW-SE
outflows can be naturally interpreted as shock-heated gas, while the
lack of velocity spread in the outflow tips (Figure \ref{copv}) can
be attributed to insufficient sensitivity.

It is expected that the outflow driving process provides energetics
to not only accelerate but also heat the (ambient) gas. Estimations
made from a simple jet-driven bow shock model predict temperature
rising with outflow velocity and distance from the driving source
\citep{hat99}, while such trends are different from the predictions
of other classes of models \citep{arc07}. \citet{hat99} further
argued that comparing with the molecular cooling mechanisms, the
heating (as a consequence of acceleration) is sufficient to maintain
outflowing gas temperature a few times higher than that of ambient
materials. Since the Hubble-like kinematic structure can also be
reproduced by the jet-driven bow shock model \citep{lee01, arc07},
all above-mentioned features of the EHV gas associated with G5.89
can be qualitatively interpreted by the jet-driven bow shock model.
We note that most, if not all, available bow shock models have
parameters typical of low-mass outflows. Models with physical
conditions more similar to outflows from high-mass young stars will
be desired to have quantitative comparisons between observational
results and model predictions. Observations with better
sensitivities and spatial resolutions are also essential to search
for the bow shock signatures in both morphology and kinematics.

\acknowledgments We thank the anonymous referee for valuable
comments. We thank all SMA staff for their help during these
observations. S.-Y. L., Y.-N. S., and H.-R. C. thank the National
Science Council of Taiwan for support this work through grants NSC
99-2112-M-001-025-MY2 and NSC 100-2112-M-007-004-MY2.

\clearpage

\begin{deluxetable}{cccccccccccc}
\singlespace
 \rotate
 \tabletypesize{\footnotesize} \tablecaption{\sc Observed CO (2$-$1) Intensities and CO (3$-$2)/(2$-$1) Line Ratios of the Five Outflow Lobes}
 \tablewidth{0pt}
 \tablehead{\multicolumn{7}{c}{BLUESHIFTED COMPONENTS} & \multicolumn{5}{c}{REDSHIFTED COMPONENTS} \\
            \colhead{LSR Outflow} & \multicolumn{2}{c}{B-NW Lobe}  & \multicolumn{2}{c}{B-N Lobe}  & \multicolumn{2}{c}{B-S Lobe} & \colhead{LSR Outflow} & \multicolumn{2}{c}{R-N Lobe}  & \multicolumn{2}{c}{R-S Lobe} \\
            \colhead{Velocity Ranges} & \colhead{2$-$1$^a$}  & \colhead{ratio$^b$} & \colhead{2$-$1$^a$}  & \colhead{ratio$^b$} & \colhead{2$-$1$^a$}  & \colhead{ratio$^b$} & \colhead{Velocity Ranges} & \colhead{2$-$1$^a$}  & \colhead{ratio$^b$} & \colhead{2$-$1$^a$}  & \colhead{ratio$^b$} \\
            \colhead{(km s$^{-1}$)} & \colhead{(K)} & \colhead{} & \colhead{(K)} & \colhead{} & \colhead{(K)} & \colhead{} & \colhead{(km s$^{-1}$)} & \colhead{(K)} & \colhead{} & \colhead{(K)} & \colhead{}  }
 \startdata
  $-$160 to $-$150   &   0.21       & 0.62     &    ---        & ---      &   ---        & ---   &   40 to 50  & ~20.08$^f$&~0.87$^f$& ~20.08$^f$ & ~0.87$^f$ \\
  $-$150 to $-$140   &   0.41       & 0.73     &    ---        & ---      &   ---        & ---   &   50 to 60  & 13.16     & 1.01    &   8.51     &  0.83 \\
  $-$140 to $-$130   &   0.40       & 0.50     &    ---        & ---      &   ---        & ---   &   60 to 70  &  6.75     & 1.10    &   2.23     &  0.56 \\
  $-$130 to $-$120   &   0.80       & 0.59     &   0.10        & 1.50     &   ---        & ---   &   70 to 80  &  3.73     & 1.05    &   ---      &   ---   \\
  $-$120 to $-$110   &   1.89       & 0.72     &   0.28        & 0.61     &   ---        & ---   &   80 to 90  &  0.72     & 0.80    &   ---      &   ---    \\
  $-$110 to $-$100   &   1.49       & ---$^c$  &   1.21        & ---$^c$  &   ---        & ---   \\
  $-$100 to $-$90    &   1.78       & 0.73     &   1.78        & 1.10     &   ---        & ---   \\
  $-$90 to $-$80     &   2.04       & 0.78     &   2.29        & 1.07     &   0.81       & 0.60  \\
  $-$80 to $-$70     &   2.84       & 0.82     &   1.92        & 1.15     &   3.70       & 0.92  \\
  $-$70 to $-$60     &   3.99       & 0.94     &   2.14        & 1.13     &   6.61       & 0.91  \\
  $-$60 to $-$50     &   6.35       & 0.94     &   2.61        & 1.19     &  10.54       & 0.87  \\
  $-$50 to $-$40     & 11.08$^d$    & ~0.87$^d$&  11.08$^d$    & ~0.87$^d$ &  14.41      & 0.84  \\
  $-$40 to $-$30     & 18.92$^e$    & ~0.84$^e$&  18.92$^e$    & ~0.84$^e$ &  ~18.92$^e$ & ~0.84$^e$  \\
\enddata
 \tablenotetext{a}{measured at the peak position of the CO (2$-$1) map}
 \tablenotetext{b}{3$-$2/2$-1$ measured at the peak position of the CO (2$-$1) map}
 \tablenotetext{c}{CO (3$-$2) blended with $^{34}$SO$_2$, which can also be discerned from Figure \ref{copv}}
 \tablenotetext{d}{including B-NW lobe and B-N lobe }
 \tablenotetext{e}{including B-NW, B-N, and B-S lobes}
 \tablenotetext{f}{including R-N lobe and R-S lobe }
\label{tb-ratio-b}
\end{deluxetable}
\end{document}